\documentclass[twocolumn,showpacs,preprintnumbers,amsmath,amssymb,prl]{revtex4-1}

\usepackage{graphicx}
\usepackage{dcolumn}
\usepackage{bm}

\begin{document}

\title{Force and flow transition in plowed granular media}
\author{Nick Gravish}
\author{Daniel I.\ Goldman}
\affiliation{School of Physics, Georgia Institute of Technology, Atlanta, Georgia 30332, USA}
\author{Paul B.\ Umbanhowar}
\affiliation{Department of Mechanical Engineering, Northwestern University, Evanston, Illinois 60208, USA}
\date{\today}

\begin{abstract}
We use plate drag to study the response of granular media to localized forcing as a function of volume fraction, $\phi$.  A bifurcation in the force and flow occurs at the onset of dilatancy, $\phi_c.$ Below $\phi_c$ rapid fluctuations in the drag force, $F_D,$ are observed. Above $\phi_c$ fluctuations in $F_D$ are periodic and increase with $\phi$. Velocity field measurements indicate that the bifurcation in $F_D$ results from the formation of stable shear bands above $\phi_c$ which are created and destroyed periodically during drag. A friction-based model captures the dynamics for $\phi>\phi_c$.
\end{abstract}

\maketitle	

Granular materials fascinate because they can act like both fluids and solids \cite{Jaeger96}. Recent work has focused on the static problem of mechanical rigidity (jamming) in which the packing density $\phi$ (the ratio of solid to occupied volume \cite{NeddermanBook}) is increased until grains crowd sufficiently to develop a finite yield stress \cite{NagelLiuRev}. Less work has explored the related process of ``unjamming" \cite{CandelierPRL09, *KerteszUnjam, *BehringerLocal, *SilbertUnjam} where initially jammed granular ensembles flow in response to forcing and where the initial packing density plays an important role: varying $\phi$ changes the local packing structure of grains which in turn affects the flow and force dynamics of the material response.

We are interested in granular media subject to localized forcing (for instance from limbs \cite{Mazouchova10, *ChenPNAS, *MaladenScience}). In general, granular systems sheared at the boundaries evolve to a steady-state $\phi_c$ \cite{SchofieldCritState}: depending on initial $\phi,$ the medium compacts ($\phi<\phi_c$) or dilates ($\phi>\phi_c$) as $\phi \rightarrow \phi_c$. In contrast, localized forcing, realized by an intruder translating through an initially homogeneous medium  \cite{CandelierPRL09, *KerteszUnjam, *BehringerLocal, Albert00} and viewed in the reference frame of the intruder, drives material only in the vicinity of the intruder toward $\phi_c$ while simultaneously advecting undisturbed media into the flowing region. The result is the repeated unjamming of ``fresh'' material and the possibility of complex spatio-temporal dynamics in and around the zone of disturbed material surrounding the intruder.

Here we drag a flat plate through initially homogeneous granular media at prepared $\phi$ to continuously drive the system away from $\phi_c$ and measure the resulting force and flow fields. As $\phi$ increases, the onset of shear dilatancy at $\phi_c$ drives a bifurcation in force dynamics and media flow. This bifurcation is governed by a complex spatiotemporal flow response associated with the stability (instability) of shear bands nucleated by the plow above (below) $\phi_c$ and their evolution in response to plow motion.  A model of shear band dynamics captures the oscillatory behavior above $\phi_c$ and suggests complex dynamics below $\phi_c.$

\begin{figure}[b]
\begin{centering}
\includegraphics[scale=0.4]{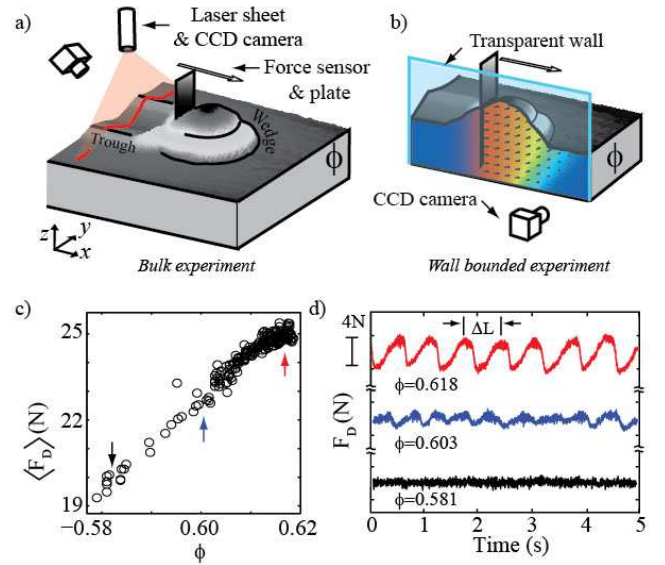}
\caption{Experimental overview:  (a) drag force, surface deformation and (b) velocity fields are measured as a function of prepared volume fraction, $\phi.$ (c) Mean drag force increases linearly with $\phi$ while (d) temporal fluctuations in $F_D$ become periodic as $\phi$ is increased (arbitrary vertical shift).}
\end{centering}
\end{figure}

{\em Experiment}--	Drag measurements, see Fig.~1(a,b), were performed in a 27$\times$86$\times$13\,cm$^3$ bed of polydisperse 256$\pm$44\,$\mu$m glass beads (Potters Industries; density $\rho=2.51$~g\,cm$^{-3}$). Similar effects to those described here were observed in other granular materials (see EPAPS \cite{EPAPS6,*EPAPS7,*EPAPS8}), including heterogeneous beach sand and poppy seeds. Air flow through the porous floor initially fluidized the medium and then a combination of air flow (below fluidization) and mechanical vibration generated the desired initial volume fraction (0.579$<\phi<$0.619). Air flow was turned off prior to testing and volume fraction determined from bed height images as $\phi = M/\rho A h,$ where $M,$ $A$ and $h$ are the bed mass, area and height respectively.  A stepper motor and linear translation stage displaced a 3.9\,cm wide by 0.3~cm thick steel plate submerged to a variable fixed depth $7.5<d<9.5$\,cm over a distance of 50\,cm at a constant speed of $v=4$\,cm\,s$^{-1}.$ An optical encoder recorded the position and a 3-axis ATI load cell mounted between plate and translation stage measured the drag forces (sampled at 200~Hz). Using laser line profilometry, we recorded the resultant surface profile and used it to quantify the change in vertical cross sectional area, $\Delta A$, normalizing by the submerged area of the plow, $A_P=$ width $\times$ depth. Profiles were measured 35\,cm from the start of drag where the profile was in a steady state. In separate drag experiments a flat plate at depth $5.5<d<6.5$~cm was positioned against a transparent wall and displaced at $v=2$\,cm\,s$^{-1}$ parallel to the wall to image grain flow. High speed video (250~fps) of the flow was recorded and analyzed in Matlab using 
image registration with a correlation time step of 0.02\,s \cite{PIV}. The near-wall setup exhibited similar force fluctuations as the bulk but with approximately half the mean force.  We remove two systematic variations in the force --- a slow increase in the baseline force during drag ($\approx 5$\% of the mean) and the decrease in depth of the constant height intruder with increasing $\phi$\,--- by defining the drag force, $F_D,$ as the raw drag with slow drift removed (3rd order polynomial fit subtracted while preserving the mean) multiplied by a depth correction factor $(\frac{d_{LP}}{d})^2$ normalized to the loose pack depth $d_{LP}.$ Separate measurements at controlled intruder depth support this normalization technique.

{\em Force bifurcation}-- The mean drag force increases approximately linearly with $\phi$ as expected due to increased bed density and average coordination number, see Fig.~1(c).  A bifurcation, however, occurs in the force fluctuations: $F_D$ at lower $\phi$ exhibits small amplitude, fast, and $\phi$ independent fluctuations, while at higher $\phi,$ slower, periodic, and larger amplitude oscillations in $F_D$ occur which grow in duration and magnitude as $\phi$ is increased, [Fig.~1(d)].

We characterize the bifurcation by measuring the standard deviation of force, $\sigma_{F_D},$ and the average plow displacement between positive slope zero crossings of the force, $\Delta L$, as a function of initial $\phi$. As $\phi$ increases we observe a sharp bifurcation in $\sigma_{F_D}$ and $\Delta L$ [Fig.~2(a)] at the compaction/dilation transition (i.e.\ $\Delta A/A_P = 0$ at $\phi_c=0.603\pm0.003$), see Fig.~2(b). As the bifurcation is approached from below, $\sigma_{F_D}$ is small and constant; above $\phi_c$ and with increasing $\phi,$ $\sigma_F$ grows linearly while $\Delta L$ increases nearly linearly 
For $\phi>\phi_c,$ $\Delta L$ is independent of plow speed ($2 < v < 8$~cm/s), see Fig.~2(a) inset, revealing it as a characteristic spatial scale and implying that the temporal dynamics of $F_D$ result from granular flow mechanisms that change with $\phi$ across the dilation transition.

\begin{figure}[htb]
\begin{centering}
\includegraphics[scale=0.35]{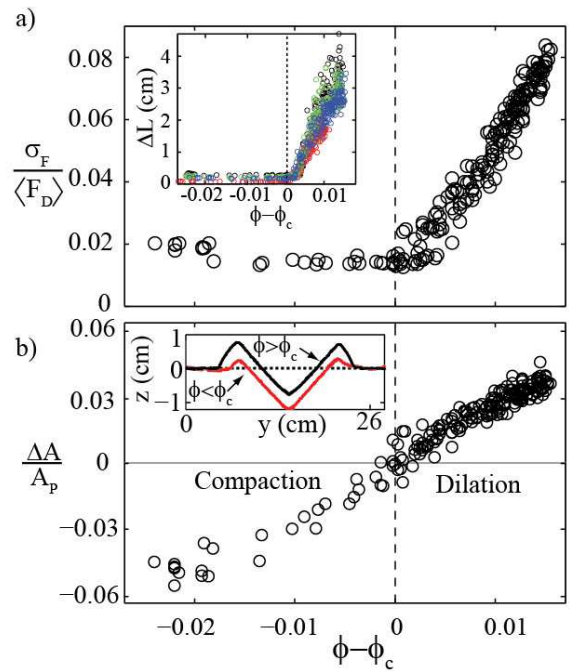}
\caption{Bifurcation in force occurs at the onset of dilation $\phi_c=0.603\pm0.003$. Small and nearly constant for low $\phi,$ (a) standard deviation of $F_D$ grows linearly above $\phi_c$ while (inset) force oscillation length scale increases nearly linearly (2, 4, 6, 8\,cm\,s$^{-1}$ in red, blue, green, and black). (b) Compacting and dilating response of plowed granular packings vs.\ initial $\phi$. (Inset) Trough cross sections at high (black) and low (red) $\phi.$ }
\end{centering}
\end{figure}

{\em Flow bifurcation}-- Direct evidence of a bifurcation in the flow is revealed by observations of the displacement of material at the surface near the plow [Fig.~3(a)]. Below $\phi_c$ surface deformation is smooth, and the boundary between upwardly moving grains and the undisturbed surface advances uniformly at a fixed distance ahead of the plow. At and above $\phi_c,$ however, the surface flow takes the form of periodic radial upwellings of grains which give the region around the plow a stepped appearance. The generation of surface ripples is correlated with fluctuations in $F_D$ and indicates that the bifurcation in $\sigma_{F_D}$ and $\Delta L$ results from a change in grain flow dynamics.

To gain insight into the nature of the bifurcation we compare the grain velocity field at low and high $\phi$ in a vertical plane perpendicular to the displacement of the plow, see Fig.~3(b) and \cite{EPAPS1, *EPAPS2}.  In both cases, flow is largely confined to a wedge-shaped region with angle $\theta$ in which grains move forward and upward in advance of the plow. At both $\phi$'s a shear band separates the flowing wedge of grains and the effectively solidified grains outside the wedge; the lower end of the shear band terminates at the bottom of the plate while the upper end reaches the surface.

\begin{figure}[htb]
\begin{centering}
\includegraphics[scale=0.24]{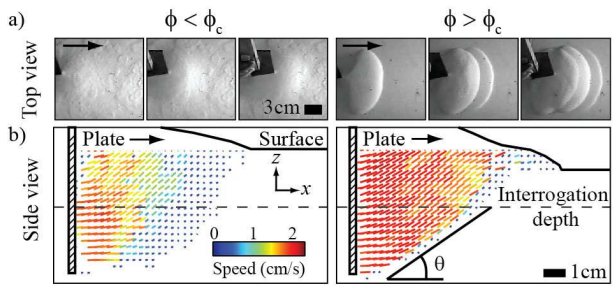}
\caption{Surface flow and velocity fields differ below $\phi_c$ (left) and above $\phi_c$ (right). (a) Surface deformation from a submerged plate shows step like features above $\phi_c.$  (b) Snapshots showing differences in velocity field structure from transparent wall experiments [Fig.~1(b)]}
\end{centering}
\end{figure}

{\em Temporal flow dynamics}-- Despite these common features, differences in spatiotemporal evolution exist between flows at low and high $\phi.$  At low $\phi,$ flow is intermittent [Fig.~4(a)]: the horizontal extent of the flowing region, $\eta,$ (measured at approximately half the intruder depth, 3.2\,cm) rapidly advances and retreats as the plow moves into new material. At high $\phi,$ the spatiotemporal evolution of the flowing region is periodic with $\eta$ remaining fixed for long periods of time until rapid, repeated fluctuations precede a jump forward to a new location. We quantify the pinning of this flow boundary as the fraction of time $\eta$ remains stationary ($\frac{d \eta}{dt}=0$) during the entire drag. As $\phi$ increases $\eta$ becomes increasingly stationary indicating that the spatial shear regions are more stable.  Above $\phi \approx \phi_c$, the shear band remains stationary for more than 50\% of the drag duration.

Shear zone localization can be understood by the process of granular shear weakening/strengthening. Below $\phi_c$ material compacts and strengthens under shear, continually frustrating the development of fixed shear zones and forcing shear to occur along a constantly changing set of failure surfaces. Above $\phi_c$ however, material dilates and weakens under shear, causing flow to localize along a fixed plane. In a plowed system, shear localization for initial $\phi > \phi_c$ holds only shortly after shear band formation: as the plow advances into undisturbed material the localized band is forced to adjust. Using the average angle of the flow boundary, $\theta,$ to characterize the orientation of the shear zone [see Fig.~3(b)], we find that below $\phi_c$ force and flow are largely uncorrelated while above $\phi_c$, $\theta$ and $F_D$ are strongly correlated with $F_D$ and $\theta$ increasing in concert, see examples in Fig.~5(a).

\begin{figure}[htb]
\begin{centering}
\includegraphics[scale=0.5]{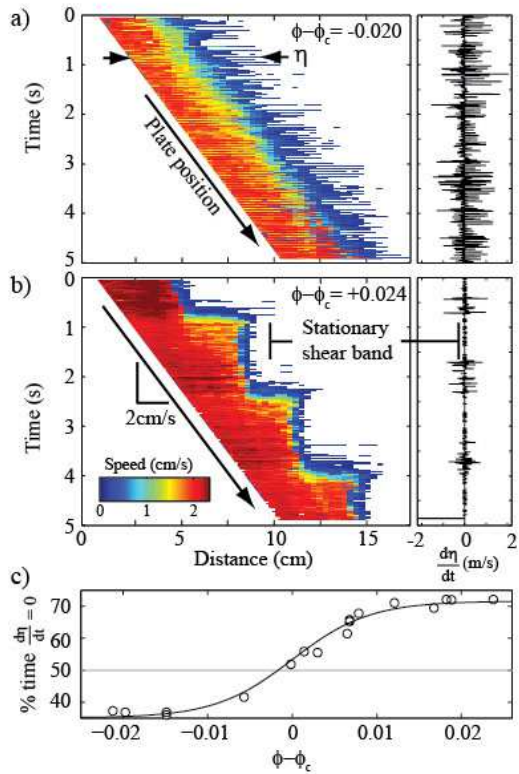}
\caption{Temporal evolution of flow differs below and above $\phi_c.$ (left) Space-time plots of flow speed along a constant depth horizontal strip show that the leading edge of the flow, $\eta,$ (a) fluctuates in loose pack and (b) is stationary between shear band formation events (four shown) in close pack. (right) Time series of $\frac{d\eta}{dt}$ highlight changes in stability. (c) Percentage of time during 5\,s drag that the flowing region is stationary ($\frac{d\eta}{dt}=0$) increases with $\phi$ and is fit to $\left[1+e^{-(\phi-\phi_c)}\right]^{-1}.$}
\end{centering}
\end{figure}

{\em Wedge flow model}-- We develop a simple model based on the correlation of $F_D$ and $\theta$ to gain insight into the dependence of $F_D$ on the flow state and $\phi.$  As the bottom of the shear zone is pinned to the lowest point of the plow, the flowing region is taken to be a triangular wedge of mass $m=\frac{\rho W d^2}{2 \tan\theta}$ sliding up a plane inclined at angle $\theta.$ Kinetic friction between the wedge and plane, $\mu(\phi),$ is assumed to increase with $\phi.$  Balancing the forces from the plow (assumed horizontal), gravity, and sliding friction, the model predicts the plowing force $F(\theta,\mu) = \frac{\rho W d^2}{2} \frac{1+\mu/\tan\theta}{1-\mu\tan\theta},$ where $W$ is the plow width [Fig.~5(c)]. The plowing force diverges at $\theta=0$ (infinite block) and $\theta = \tan^{-1}\left(\mu^{-1}\right)$ and is minimum at $\theta_{min} = \tan^{-1}\left(\sqrt{\mu^2+1)}-\mu\right)$.  $\theta_{min}$ decreases and $F(\theta_{min})$ increases with increasing $\mu (\phi).$

The model predicts that an initially jammed and homogeneous material with volume fraction $\phi_0$ shears at an angle $\theta_{min}$ when the force reaches $F_{min}(\mu),$ see Fig.~5(b,c). For $\phi_0 > \phi_c,$ shearing along the slip plane dilates and weakens the material locally, causing $\mu(\phi_0)$ to decrease to $\mu(\phi_c)$ which reduces $F_D$ and slightly increases $\theta,$ see path A in Fig.~5(c).  As the plow advances the angle of the weakened shear zone gradually increases along $F[\theta,\mu(\phi_c)],$ indicated by path B. The shear band remains fixed at the surface (causing $\theta$ to increase) instead of advancing with a constant angle because, evidently, less high $\phi$ material needs to be broken in the former case. With increasing $\theta,$ the force to push the wedge $\theta_{max}$ eventually equals the force required to break the stronger material at $\phi_0>\phi_c,$ i.e.\ $F[\theta_{max},\mu(\phi_c)] = F[\theta_{min},\mu(\phi_0)].$ At this point a new shear zone forms in front of the plow, decreasing $\theta$ to $\theta_{min}(\phi_0)$ and the cycle repeats (path C).

\begin{figure}[htb]
\begin{centering}
\includegraphics[scale=0.5]{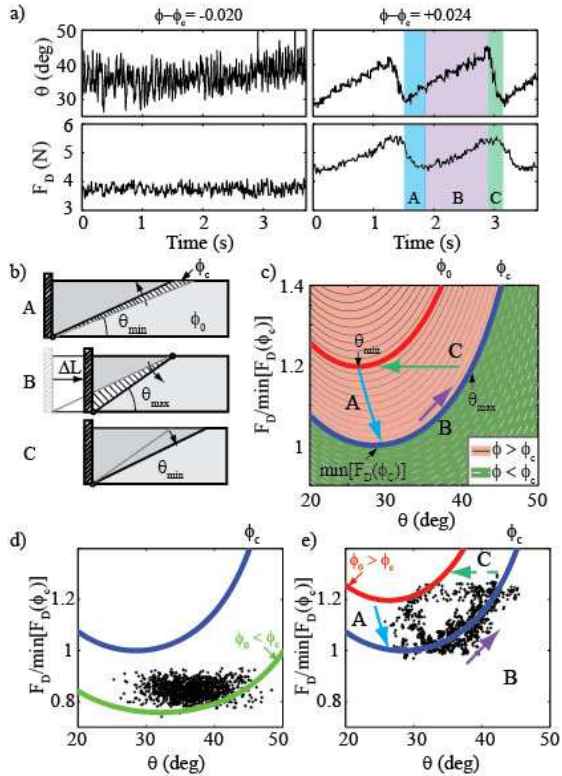}
\caption{Force dynamics and wedge model of granular flow. (a) Wedge flow angle $\theta$ and $F_D$ are correlated above $\phi_c.$ (b) Sketch of wedge flow model dynamics above $\phi_c$ with arrows indicating motion of shear band for the three phases A,B,C also shown in (a,c,e).  (c) Dynamics predicted by model (arrows) showing family of ($\theta,F_D$) curves for varying $\mu(\phi)$ and transitions between $\mu(\phi)$ curves corresponding to $\phi_0$ (red curve) and $\phi_c$ (blue curve), see text. Comparison of model ($\theta,F_D$) curves and experiment: (d) for $\phi<\phi_c$ experimental data are bounded by model curves of arbitrary $\mu<\mu(\phi_c);$ (e) for $\phi>\phi_c$ model shows good agreement with experiment.}
\end{centering}
\end{figure}

The dynamics of the experimental data for $\phi>\phi_c$ are captured by the model as seen in $(\theta,F(\theta))$ space [Fig.~5(e)] and \cite{EPAPS3,*EPAPS4,*EPAPS5}. Here the system evolves between two curves [$\mu = 0.65$ (blue) and $\mu = 0.78$ (red)] and the transitions between these states qualitatively match the predictions of our model. The model predicts the increase in force fluctuations with increasing $\phi$ above $\phi_c$. However for $\phi < \phi_c$ the evolution of $F_D$ with $\theta$ is more complicated because of the lack of stability of the sheared regions through shear strengthening. As we assume a feedback process exists between the flow dynamics and $F_D$, strengthening of the material during shear results in the relocation of the shear band to a weaker region in the surrounding bulk. With the continuous advection of undisturbed weaker material and the heterogeneity in $\phi$ in the previously sheared region, predicting the shear process below $\phi_c$ is more challenging.

In summary we observe a flow and force bifurcation in localized granular drag which occurs at the onset of dilatancy. Localized forcing in granular media induces a heterogeneous $\phi$ distribution in the region of flow and represents a new way to test theories of granular flow \cite{GDRMiDi, *VanHeckeREV}. Previous studies of drag in granular media \cite{Albert00} were performed at slow speed to observe stick-slip fluctuations over distances on the order of a grain diameter; the effects of varying initial $\phi$ were not investigated. Our observations of a bifurcation in $\phi$ support the findings of Schr\"{o}ter et al. \cite{SchroterPhase07} who reported a transition in dynamics at $\phi=0.598$, and our measured $\phi_c$ is similar to that reported ($\phi_c=0.591$) in recent impact experiments \cite{Impact}. It is known that $\phi_c$ is a function of the applied stresses and cohesive/frictional interactions between the grains \cite{Fazekas04, SchofieldCritState} so we do not place importance on the specific value here.

\begin{acknowledgments}
We thank Harry Swinney and Mathias Schr\"{o}ter for stimulating discussion, and Andrei Savu for apparatus fabrication. This work was supported by the Burroughs Wellcome Fund and the ARL MAST CTA under cooperative agreement number W911NF-08-2-0004.
\end{acknowledgments}

\bibliographystyle{apsrev4-1}

\begin{thebibliography}{10}%
\makeatletter
\providecommand \@ifxundefined [1]{%
 \ifx #1\undefined \expandafter \@firstoftwo
 \else \expandafter \@secondoftwo
\fi
}%
\providecommand \@ifnum [1]{%
 \ifnum #1\expandafter \@firstoftwo
 \else \expandafter \@secondoftwo
\fi
}%
\providecommand \enquote [1]{``#1''}%
\providecommand \bibnamefont  [1]{#1}%
\providecommand \bibfnamefont [1]{#1}%
\providecommand \citenamefont [1]{#1}%
\providecommand\href[0]{\@sanitize\@href}%
\providecommand\@href[1]{\endgroup\@@startlink{#1}\endgroup\@@href}%
\providecommand\@@href[1]{#1\@@endlink}%
\providecommand \@sanitize [0]{\begingroup\catcode`\&12\catcode`\#12\relax}%
\@ifxundefined \pdfoutput {\@firstoftwo}{%
 \@ifnum{\z@=\pdfoutput}{\@firstoftwo}{\@secondoftwo}%
}{%
 \providecommand\@@startlink[1]{\leavevmode}%
 \providecommand\@@endlink[0]{}%
}{%
 \providecommand\@@startlink[1]{%
  \leavevmode
  \pdfstartlink
   attr{/Border[0 0 1 ]/H/I/C[0 1 1]}%
   user{/Subtype/Link/A<</Type/Action/S/URI/URI(#1)>>}%
  \relax
 }%
 \providecommand\@@endlink[0]{\pdfendlink}%
}%
\providecommand \url  [0]{\begingroup\@sanitize \@url }%
\providecommand \@url [1]{\endgroup\@href {#1}{\urlprefix}}%
\providecommand \urlprefix [0]{URL }%
\providecommand \Eprint[0]{\href }%
\@ifxundefined \urlstyle {%
  \providecommand \doi [1]{doi:\discretionary{}{}{}#1}%
}{%
  \providecommand \doi [0]{doi:\discretionary{}{}{}\begingroup
  \urlstyle{rm}\Url }%
}%
\providecommand \doibase [0]{http://dx.doi.org/}%
\providecommand \Doi[1]{\href{\doibase#1}}%
\providecommand \bibAnnote [3]{%
  \BibitemShut{#1}%
  \begin{quotation}\noindent
    \textsc{Key:}\ #2\\\textsc{Annotation:}\ #3%
  \end{quotation}%
}%
\providecommand \bibAnnoteFile [2]{%
  \IfFileExists{#2}{\bibAnnote {#1} {#2} {\input{#2}}}{}%
}%
\providecommand \typeout [0]{\immediate \write \m@ne }%
\providecommand \selectlanguage [0]{\@gobble}%
\providecommand \bibinfo [0]{\@secondoftwo}%
\providecommand \bibfield [0]{\@secondoftwo}%
\providecommand \translation [1]{[#1]}%
\providecommand \BibitemOpen[0]{}%
\providecommand \bibitemStop [0]{}%
\providecommand \bibitemNoStop [0]{.\EOS\space}%
\providecommand \EOS [0]{\spacefactor3000\relax}%
\providecommand \BibitemShut [1]{\csname bibitem#1\endcsname}%
\bibitem{Jaeger96}%
  \BibitemOpen
  \bibfield{author}{B
  \bibinfo {author} {\bibfnamefont{H.~M.}\ \bibnamefont{Jaeger}}, \bibinfo
  {author} {\bibfnamefont{S.~R.}\ \bibnamefont{Nagel}},\ and\ \bibinfo {author}
  {\bibfnamefont{R.}~\bibnamefont{Behringer}},\ }%
  \bibfield{journal}{%
  \bibinfo {journal} {Rev. Mod. Phys.}\ }%
  \textbf{\bibinfo {volume} {68}} (\bibinfo {year} {1996})%
  \bibAnnoteFile{NoStop}{Jaeger96}%
\bibitem{NeddermanBook}%
  \BibitemOpen
  \bibfield{author}{%
  \bibinfo {author} {\bibfnamefont{R.}~\bibnamefont{Nedderman}},\ }%
  \emph{\bibinfo {title} {Statics and Kinematics of Granular Materials}}\
  (\bibinfo {publisher} {Cambridge University Press},\ \bibinfo {year} {1992})%
  \bibAnnoteFile{NoStop}{NeddermanBook}%
\bibitem{NagelLiuRev}%
  \BibitemOpen
  \bibfield{author}{%
  \bibinfo {author} {\bibfnamefont{A.}~\bibnamefont{Liu}}\ and\ \bibinfo
  {author} {\bibfnamefont{S.}~\bibnamefont{Nagel}},\ }%
  \bibfield{journal}{%
  \bibinfo {journal} {Annual Review of Condensed Matter Physics}\ }%
  \textbf{\bibinfo {volume} {1}} (\bibinfo {year} {2010})%
  \bibAnnoteFile{NoStop}{NagelLiuRev}%
\bibitem{CandelierPRL09}%
  \BibitemOpen
  \bibfield{author}{%
  \bibinfo {author} {\bibfnamefont{R.}~\bibnamefont{Candelier}}\ and\ \bibinfo
  {author} {\bibfnamefont{O.}~\bibnamefont{Dauchot}},\ }%
  \bibfield{journal}{%
  \bibinfo {journal} {Phys. Rev. Lett.}\ }%
  \textbf{\bibinfo {volume} {103}},\ \bibinfo {pages} {128001} (\bibinfo {year}
  {2009})%
  \bibAnnoteFile{NoStop}{CandelierPRL09}%
\bibitem{KerteszUnjam}%
  \BibitemOpen
  \bibfield{author}{%
  \bibinfo {author} {\bibfnamefont{M.}~\bibnamefont{Shaebani}}, \bibinfo
  {author} {\bibfnamefont{T.}~\bibnamefont{Unger}},\ and\ \bibinfo {author}
  {\bibfnamefont{J.}~\bibnamefont{Kert\'{e}sz}},\ }%
  \bibfield{journal}{%
  \bibinfo {journal} {Phys. Rev. E}\ }%
  \textbf{\bibinfo {volume} {78}} (\bibinfo {year} {2008})%
  \bibAnnoteFile{NoStop}{KerteszUnjam}%
\bibitem{BehringerLocal}%
  \BibitemOpen
  \bibfield{author}{%
  \bibinfo {author} {\bibfnamefont{J.~Geng}\ \bibnamefont{et~al.}},\ }%
  \bibfield{journal}{%
  \bibinfo {journal} {Phys. Rev. Lett.}\ }%
  \textbf{\bibinfo {volume} {87}} (\bibinfo {year} {2001})%
  \bibAnnoteFile{NoStop}{BehringerLocal}%
\bibitem{SilbertUnjam}%
  \BibitemOpen
  \bibfield{author}{%
  \bibinfo {author} {\bibfnamefont{L.}~\bibnamefont{Silbert}}, \bibinfo
  {author} {\bibfnamefont{A.}~\bibnamefont{Liu}},\ and\ \bibinfo {author}
  {\bibfnamefont{S.}~\bibnamefont{Nagel}},\ }%
  \bibfield{journal}{%
  \bibinfo {journal} {Phys. Rev. E}\ }%
  \textbf{\bibinfo {volume} {73}} (\bibinfo {year} {2006})%
  \bibAnnoteFile{NoStop}{SilbertUnjam}%
\bibitem{Impact}%
  \BibitemOpen
  \bibfield{author}{%
  \bibinfo {author} {\bibfnamefont{P.}~\bibnamefont{Umbanhowar}},\ and\ \bibinfo
  {author} {\bibfnamefont{D.}~\bibnamefont{Goldman}},\ }%
  \bibfield{journal}{%
  \bibinfo {journal} {Phys. Rev. E}\ }%
  \textbf{\bibinfo {volume} {}} (\bibinfo {year} {2010})%
  \bibAnnoteFile{NoStop}{SilbertUnjam}%
\bibitem{Mazouchova10}%
  \BibitemOpen
  \bibfield{author}{%
  \bibinfo {author} {\bibfnamefont{N.}~\bibnamefont{Mazouchova}}
  \emph{et~al.},\ }%
  \bibfield{journal}{%
  \bibinfo {journal} {Biology Letters}}%
   (\bibinfo {year} {2010})%
  \bibAnnoteFile{NoStop}{Mazouchova10}%
\bibitem{ChenPNAS}%
  \BibitemOpen
  \bibfield{author}{%
  \bibinfo {author} {\bibfnamefont{C.}~\bibnamefont{Li}} \emph{et~al.},\ }%
  \bibfield{journal}{%
  \bibinfo {journal} {PNAS}\ }%
  \textbf{\bibinfo {volume} {106}} (\bibinfo {year} {2009})%
  \bibAnnoteFile{NoStop}{ChenPNAS}%
\bibitem{MaladenScience}%
  \BibitemOpen
  \bibfield{author}{%
  \bibinfo {author} {\bibfnamefont{R.}~\bibnamefont{Maladen}} \emph{et~al.},\
  }%
  \bibfield{journal}{%
  \bibinfo {journal} {Science}\ }%
  \textbf{\bibinfo {volume} {325}} (\bibinfo {year} {2009})%
  \bibAnnoteFile{NoStop}{MaladenScience}%
\bibitem{SchofieldCritState}%
  \BibitemOpen
  \bibfield{author}{%
  \bibinfo {author} {\bibfnamefont{A.}~\bibnamefont{Schofield}}\ and\ \bibinfo
  {author} {\bibfnamefont{C.}~\bibnamefont{Wroth}},\ }%
  \emph{\bibinfo {title} {Critical State Soil Mechanics}}\ (\bibinfo
  {publisher} {McGraw-Hill},\ \bibinfo {year} {1968})%
  \bibAnnoteFile{NoStop}{SchofieldCritState}%
\bibitem{Albert00}%
  \BibitemOpen
  \bibfield{author}{%
  \bibinfo {author} {\bibfnamefont{I.}~\bibnamefont{Albert}}, \emph{et~al.},\
  }%
  \bibfield{journal}{%
  \bibinfo {journal} {Phys. Rev. Lett.}\ }%
  \textbf{\bibinfo {volume} {84}} (\bibinfo {year} {2000})%
  \bibAnnoteFile{NoStop}{Albert00}%
\bibitem{PIV}%
  \BibitemOpen
  \bibfield{author}{%
  \bibinfo {author} {\bibfnamefont{M.}~\bibnamefont{Guizar-Sicairos}}, \bibinfo
  {author} {\bibfnamefont{S.}~\bibnamefont{Thurman}},\ and\ \bibinfo {author}
  {\bibfnamefont{J.}~\bibnamefont{Fienup}},\ }%
  \bibfield{journal}{%
  \bibinfo {journal} {Opt. Lett.}\ }%
  \textbf{\bibinfo {volume} {33}},\ \bibinfo {pages} {156} (\bibinfo {year}
  {2008})%
  \bibAnnoteFile{NoStop}{PIV}%
\bibitem{EPAPS1}%
  \BibitemOpen
  \bibinfo {author} {\bibnamefont{EPAPS:CP-Two-Color-xvd}}%
  \bibAnnoteFile{NoStop}{EPAPS1}%
\bibitem{EPAPS2}%
  \BibitemOpen
\bibfield{author}{%
    }%
  \bibinfo {author} {\bibnamefont{EPAPS:LP-Two-Color-xvd}}%
  \bibAnnoteFile{NoStop}{EPAPS2}%
\bibitem{EPAPS3}%
  \BibitemOpen
\bibfield{author}{%
    }%
  \bibinfo {author} {\bibnamefont{EPAPS:Wedge-dynamics}}%
  \bibAnnoteFile{NoStop}{EPAPS3}%
\bibitem{EPAPS4}%
  \BibitemOpen
\bibfield{author}{%
    }%
  \bibinfo {author} {\bibnamefont{dynamics CP}}%
  \bibAnnoteFile{NoStop}{EPAPS4}%
\bibitem{EPAPS5}%
  \BibitemOpen
\bibfield{author}{%
    }%
  \bibinfo {author} {\bibnamefont{dynamics LP}}%
  \bibAnnoteFile{NoStop}{EPAPS5}%
\bibitem{EPAPS6}%
  \BibitemOpen
\bibfield{author}{%
    }%
  \bibinfo {author} {\bibfnamefont{EPAPS:CP-Jekyll}}%
  \bibAnnoteFile{NoStop}{EPAPS6}%
\bibitem{EPAPS7}%
  \BibitemOpen
\bibfield{author}{%
    }%
  \bibinfo {author} {\bibnamefont{LP-Jekyll}}%
  \bibAnnoteFile{NoStop}{EPAPS7}%
\bibitem{EPAPS8}%
  \BibitemOpen
\bibfield{author}{%
    }%
  \bibinfo {author} {\bibnamefont{Jekyll Sand}}%
  \bibAnnoteFile{NoStop}{EPAPS8}%
\bibitem{SchroterPhase07}%
  \BibitemOpen
\bibfield{author}{%
    }%
  \bibfield{author}{%
  \bibinfo {author} {\bibfnamefont{M.}~\bibnamefont{Schr\"{o}ter}}
  \emph{et~al.},\ }%
  \bibfield{journal}{%
  \bibinfo {journal} {EPL}\ }%
  \textbf{\bibinfo {volume} {78}} (\bibinfo {year} {2007})%
  \bibAnnoteFile{NoStop}{SchroterPhase07}%
\bibitem{Fazekas04}%
  \BibitemOpen
  \bibfield{author}{%
  \bibinfo {author} {\bibfnamefont{S.}~\bibnamefont{Fazekas}}, \bibinfo
  {author} {\bibfnamefont{J.}~\bibnamefont{Torok}},\ and\ \bibinfo {author}
  {\bibfnamefont{J.}~\bibnamefont{Kert\'{e}sz}},\ }%
  \bibfield{journal}{%
  \bibinfo {journal} {Phys. Rev. E}\ }%
  \textbf{\bibinfo {volume} {75}} (\bibinfo {year} {2007})%
  \bibAnnoteFile{NoStop}{Fazekas04}%
\bibitem{GDRMiDi}%
  \BibitemOpen
  \bibfield{author}{%
  \bibinfo {author} {\bibfnamefont{G.}~\bibnamefont{MiDi}},\ }%
  \bibfield{journal}{%
  \bibinfo {journal} {Eur. Phys. J. E.}\ }%
  \textbf{\bibinfo {volume} {14}} (\bibinfo {year} {2004})%
  \bibAnnoteFile{NoStop}{GDRMiDi}%
\bibitem{VanHeckeREV}%
  \BibitemOpen
  \bibfield{author}{%
  \bibinfo {author} {\bibfnamefont{P.}~\bibnamefont{Schall}}\ and\ \bibinfo
  {author} {\bibfnamefont{M.}~\bibnamefont{van Hecke}},\ }%
  \bibfield{journal}{%
  \bibinfo {journal} {Ann. Rev. of F. Mech.}\ }%
  \textbf{\bibinfo {volume} {42}} (\bibinfo {year} {2010})%
  \bibAnnoteFile{NoStop}{VanHeckeREV}%
\end{thebibliography}

%

\end{document}